\documentclass{elsart3}

\usepackage{graphics}
\usepackage{graphicx}

\usepackage{amssymb}

\begin{document}

\begin{frontmatter}

\title{CMS Barrel Pixel Detector Overview}

\author[psi]{H. Chr. K\"astli\corauthref{cor}},
\author[psi]{W. Bertl},
\author[psi]{W. Erdmann},
\author[psi]{K. Gabathuler}
\author[unizh,psi]{Ch. H\"ormann},
\author[psi]{R. Horisberger},
\author[psi]{S. K\"onig},
\author[psi]{D. Kotlinski},
\author[eth]{B. Meier},
\author[unizh]{P. Robmann},
\author[psi]{T. Rohe},
\author[eth]{S. Streuli}

\address[psi]{Paul Scherrer Institut, 5232 Villigen PSI, Switzerland}
\address[unizh]{Physik-Institut der Univerit\"at Z\"urich, 8057 Z\"urich, Switzerland}
\address[eth]{Institut f\"ur Teilchenphysik, ETH Z\"urich, 8093 Z\"urich, Switzerland}
\corauth[cor]{Corresponding author.\\ {\em Email: hans-christian.kaestli@psi.ch}}

\begin{abstract}
The pixel detector is the innermost tracking device of the CMS experiment at the LHC. It is built from two independent sub devices, the pixel barrel and the end disks. The barrel consists of three concentric layers around the beam pipe with mean radii of 4.4, 7.3 and 10.2 cm. There are two end disks on each side of the interaction point at $\pm$34.5 cm and $\pm$46.5 cm\\
This article gives an overview of the pixel barrel detector, its mechanical support structure, electronics components, services  and its expected performance.
\end{abstract}

\end{frontmatter}

\begin{figure*}[ht]
\begin{center}
\includegraphics[width=12cm]{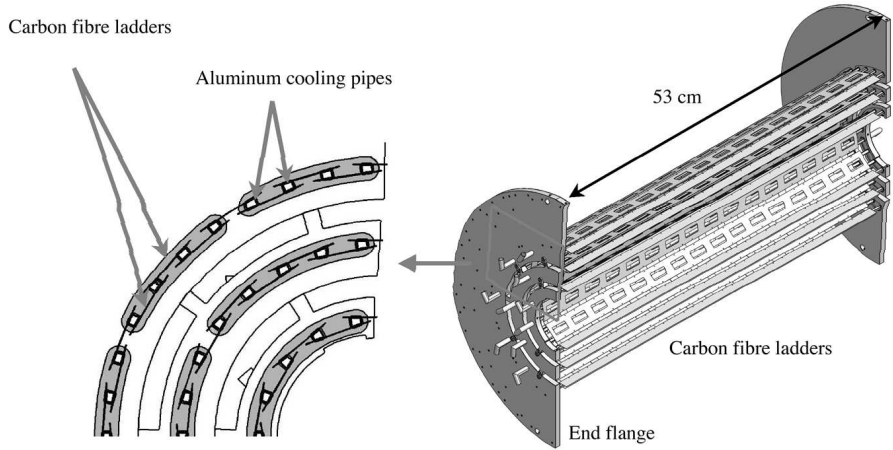}
\end{center}
\caption{Drawing of half of the barrel support structure (right) and detailed view of the geometric layout (left)}
\label{fig:mechanics}
\end{figure*}

\section{Introduction}

Tracking and vertexing in a high multiplicity environment such as the LHC is a challenging task. In order to get a good secondary vertex resolution, one needs measurements as close to the interaction point as possible. To have acceptably low occupancies one needs a very fine segmented detector. The CMS experiment  uses a hybrid pixel detector with $100\times 150 \mu$m$^2$ pixel size. The barrel part has three layers at mean radii of 4.4, 7.3 and 10.2 cm. In the innermost layer the occupancy reaches $6\cdot 10^{-4}$ at the full LHC luminosity of $10^{34}$s$^{-1}$cm$^{-2}$.\\
The main purpose of the pixel detector is the reconstruction of secondary vertices (mainly for b and $\tau$ physics) and the generation of track seeds for the reconstruction in the full tracker. Also tracks reconstructed with the pixel detector alone is the only available tracking information at the first stage of the higher level trigger (HLT). Here speed is more important than accuracy or efficiency. As pointed out in \cite{PTDRI}, pixel-only tracks can be reconstructed in less than 20ms (110ms) per event for regional (global) track finder. The $p_T$ resolution is $\frac{\sigma (p_T)}{p_T}=0.055+ 0.017\cdot p_T$[GeV/c] and the impact parameter can be measured with $\sigma(IP)=80\mu$m for $p_T>7$GeV/c rising to $90\mu$m for $p_T=4$ GeV/c. The component of the vertex position along the beam line is measured at the HLT. The resolution depends on the physics channel (number and momentum of tracks) and is typically $\lessapprox 50 \mu$m with an efficiency of $>94\%$.

\section{Mechanical structure}

\begin{figure}[h]
\begin{center}
\includegraphics[width=7.5cm]{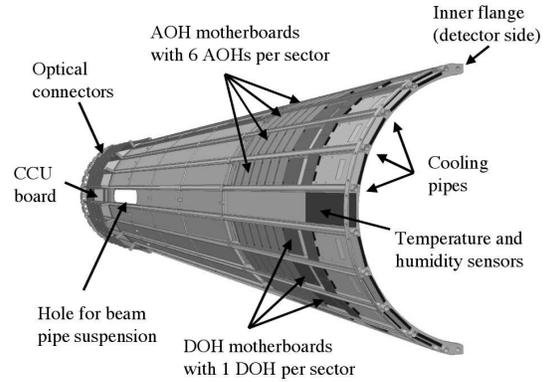}
\end{center}
\caption{Drawing of a supply tube half-cylinder. The electronics components are described in section 4}
\label{fig:supply}
\end{figure}

The smallest independent unit of the barrel support structure is a half-shell. It is made out of 53cm long and 0.25 mm thin carbon fiber ladders glued to aluminum cooling pipes with 0.3 mm wall thickness. To reach full spatial coverage ladders are mounted with overlap on alternating sides of the cooling tubes. This is shown in figure \ref{fig:mechanics} on the left. Since the ladders are not tilted, charge sharing is enhanced between pixels due to Lorentz drift in the 4T magnetic field and hence the spatial resolution through analog signal interpolation is increased. \\
Eight modules are screwed onto each ladder using thermal grease to improve the heat transfer from the electronics components to the cooling tubes. This allows an easy removal of the mounted modules. \\
Three half-shells are mounted together at the end flange building up half of the barrel detector. The left and right half barrels are mechanically separated. A six layer PCB is mounted on the end flange. It distributes power and signals to the individual modules. The PCB is equipped with LCDS (Low Current Differential Signal) driver chips and is otherwise passive. About 10cm long cables and cooling tubes connect the barrel support structure to the service half-cylinders. A sketch of the later is shown in figure \ref{fig:supply}.    

\section{Modules}

\begin{figure}
\begin{center}
\includegraphics[width=7.5cm]{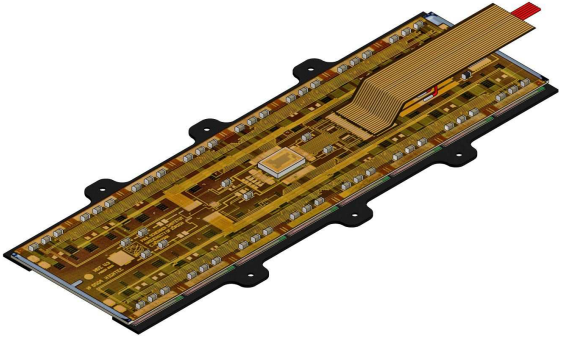}
\end{center}
\caption{Drawing of a full barrel module}
\label{fig:module}
\end{figure}

The barrel layers are made out of two types of sensor modules. Each layer has 32 half modules at the edges of the half-shells and 128, 224 and 320 full modules for the three layers respectively. A sketch of a module is shown in figure \ref{fig:module}. 16 (8) readout chips (ROCs) are bump bonded to the full (half) sensors. A dedicated In bump bonding process has been developed at PSI \cite{bumpbonding} and the whole production is done in house. A three layer flexible high density interconnect (HDI) is glued on top of the sensor and wire bonded to the ROCs and to the token bit manager chip (TBM) \cite{ed}. It is then glued to Si-Nitride base stripes which give mechanical stability and reduces stress to the bump bonds due to an excellent CTE match to Si. Details about the module assembly are described in these proceedings (\cite{stefan}).

\subsection{Sensor}

The active area of a full (half) module sensor is 64.8$\times$10.6 mm$^2$ (64.8$\times$ 5.3mm$^2$). It is a double sided processed "n+ on n" design. This is a pixelated high dose n implantation in a lightly n-doped bulk material. The backside is p-doped forming the pn junction. The inter-pixel isolation is done with a moderated p-spray technique. There are mainly two reasons for this choice: (i) after irradiation, the bulk material will be type inverted. Depletion will then begin at the structured n-side, allowing the device to be operated in a partially depleted mode and thus at a lower bias voltage. (ii) A multiple guard ring structure on the p-side controls the voltage drop from the large negative bias voltage towards the sensor edge and allows to put all sensor edges to ground potential. This enables a safe operation at very high bias voltages up to 600V.

\subsection{Readout chip}

The readout chip (ROC) is fabricated in a commercial $0.25 \mu$m 5 metal layer CMOS process. It's main purposes are
\begin{itemize}
\item amplification and buffering of the charge signal from the sensor 
\item zero suppression in the pixel unit cell. Only signals above a certain threshold will be read out. This threshold can be adjusted individually for each pixel by means of four trim bits. The trim bits have capacitive protection against single event upset (SEU), which have shown to reduce SEU by 2 orders of magnitude \cite{hadi}. A typical threshold dispersion after trimming at T=-10$^\circ$C is 90 electron equivalents with a noise of 170 electrons.  
\item level 1 trigger verification. Hit information without a corresponding L1 trigger are abandoned.
\item sending hit information and configuration data out to the TBM chip
\item adjusting various voltage levels and offsets in order to compensate for chip-to-chip variations in the CMOS device parameters 
\end{itemize}
The ROC has 6 on chip voltage regulators. This makes it possible to operate the detector without voltage regulator boards inside the CMS tracker volume, improves AC power rejection and strongly reduces intermodule cross-talk. More details about the ROC can be found in \cite{hadi}.

\begin{figure*}[ht]
\begin{center}
\includegraphics[width=10.5cm]{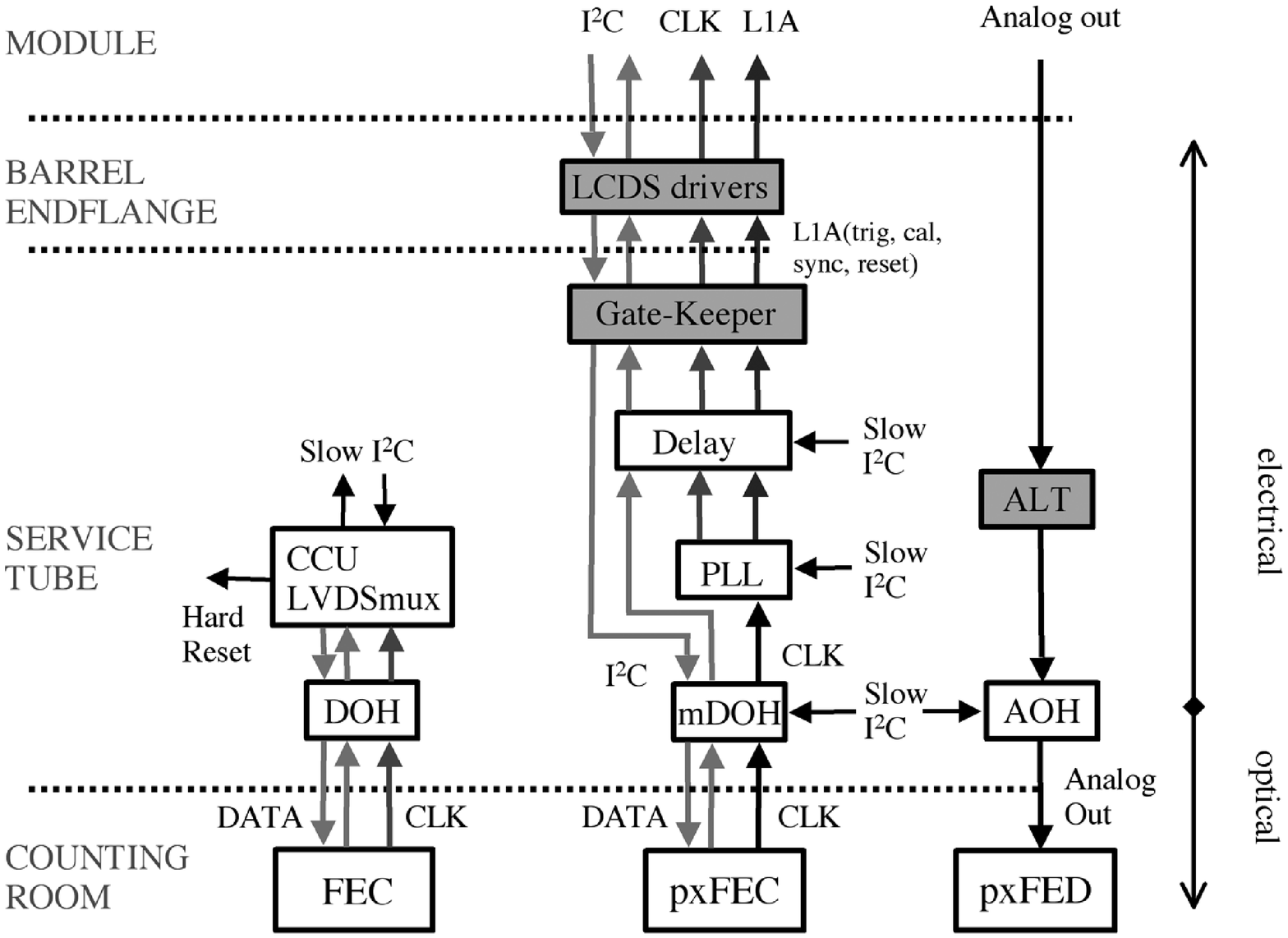}
\end{center}
\caption{Schematic view of the signal flow for the barrel pixel detector}
\label{fig:signal}
\end{figure*}

\subsection{Module readout}
The module is readout in a daisy chained scheme. It is controlled by the token bit manager (TBM) chip. For each level 1 trigger the TBM generates a token bit which controls 8 or 16 ROCs. One ROC at a time sends its hit information through a LVDS (low voltage differential signal) to the TBM chip where they are amplified and converted into a LCDS (low current differential signal). The later has been chosen because of its low voltage swing, in order to reduce cross-talk on the kapton signal cable to the supply tube.\\ 
There is one analog readout link running at 40MHz for each half module and each full module in the outermost layer and 2 such links for each full module in the two inner layers. In order to reduce readout time, the pixel address is sent in a digital coded analog signal, i.e. there exist 6 discrete address levels. The amplified sensor signal is sent purely analog. Each hit pixel needs 6 clock cycles or 150ns to be read out. The address level separation has been measured with the entire system and was found to be $>$30 times the RMS of the level width. Thus, reliable address decoding is ensured.

\section{Detector control and readout}

Figure \ref{fig:signal} shows a simplified diagram of the signal flow for the barrel pixel detector. The signal path in the middle shows the fast detector control link. The pixel front end controller (pFEC) sends a clock signal with encoded fast control signals (L1 trigger, reset, internal calibration and resynchronization signal) to the detector. The pFEC also downloads configuration data at run start and if necessary during runs via the industrial I$^2$C protocol, modified to run at a higher clock rate of 40MHz. All connections between service tube and counting room are optical links. The DOH (digital opto hybrid) does the conversion between optical and electrical signals. A PLL chip regenerates the fast signals and restores the clock. A delay chip can adjust the phase for each signal separately. The GateKeeper chip blocks idle signals needed for automatic gain adjustment of the optical links from being transmitted to the module. Finally there is a LVDS to LCDS level converter chip on the end flange print.\\
The path on the right shows the readout part. The TBM drives the analog signal all the way to the service tube. A level translator chip (ALT) generates signal levels acceptable to the laser driver chips of the analog optical link. Driver and optical transmitter are located on the analog opto hybrid (AOH). The signal is received in the counting room by a VME front end driver unit (pxFED) \cite{hephy}. It converts back to electrical signals, does a baseline correction, determines all pixel addresses and digitizes the analog pulse height. It then passes this information to the CMS DAQ system.\\
Some of these components need to be programmed. This happens through a dedicated digital link via a standard I$^2$C protocol. The boxes in gray show pixel specific full custom auxiliary ASICs developed at PSI.

\section{Performance}

\begin{figure}
\begin{center}
\includegraphics[width=7cm]{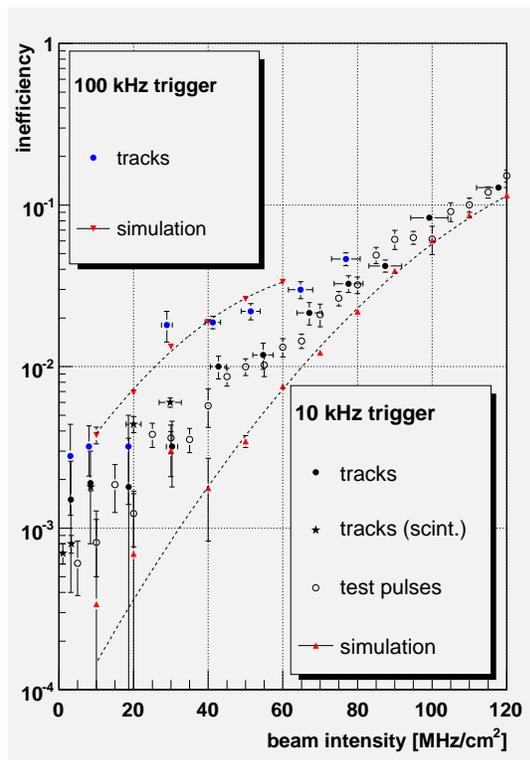}
\end{center}
\caption{Measured efficiency in a pion beam at PSI as a function of particle fluence for two different trigger rates. For comparison simulation results are shown (triangles).}
\label{fig:effi}
\end{figure}

Barrel modules have been tested in a high rate 300 MeV/c pion beam at PSI. Figure \ref{fig:effi} shows the detection efficiency as a function of particle fluence for trigger rates of 10 and 100kHz. Three different methods have been used. For low trigger rates efficiencies are shown for tracks reconstructed with a beam telescope or using small scintillators only. The third method measures the efficiency for internal calibration signals which are superimposed on top of the beam induced data traffic. For comparison, triangles show the efficiencies as expected from simulations. At high trigger rates the data losses are dominated by readout related dead times. The agreement between data and simulation is excellent here. For low trigger rates the measurement shows $\approx 0.5\%$ higher inefficiency than expected from simulations. This can be understood by radiation damage of the sensor under test. \\
Resolution studies have been made in \cite{dorokhov}. Using pion beam test data together with simulations, one expects a resolution in $r\Phi$ between 10 and 20 $\mu$m for sensors irradiated up to $6.7\cdot 10^14 $n$_{eq}/$cm$^2$, which corresponds to the first 4 years of operation in the innermost layer. The z resolution depends on the azimuthal angle $\Theta$ and only weakly on the irradiation dose. For $\Theta<60^\circ$ the resolution is $<20 \mu$m, increasing to 40 $\mu$m at $\Theta=90^\circ$.

\section{Conclusion and outlook}
An overview of the CMS barrel pixel detector has been given. The design is finalized and all system components have been tested extensively. The device is still in the construction phase and will be delivered to CMS in summer 2007. First proton collisions are expected by the end of 2007 at the injection energy of 450GeV per beam and at the full energy of 7TeV in spring 2008.


\begin{thebibliography}{00}

\bibitem{TDR} The CMS collaboration, CMS Tracker Technical Design Report, CERN/LHCC 98-6
\bibitem{PTDRI} The CMS collaboration, CMS Physics Technical Design Report, Vol I: Detector performance and Software, CERN/LHCC 06-001
\bibitem{bumpbonding} Ch. Broennimann, Development of an Indium bump bond process for silicon pixel detectors at PSI, Nucl. Instrum. Methods A 565 (2006) 303
\bibitem{stefan} S. K\"onig, Building CMS Pixel Barrel Detector Modules, These proceedings
\bibitem{hadi} H. C. K\"astli, Design and performance of the CMS pixel detector readout chip, Nucl. Instrum. Methods A 565 (2006) 188
\bibitem{danek} D. Kotlinski, The Control and Readout System of the CMS Pixel Barrel Detector, Nucl. Instrum. Methods A 565 (2006) 73
\bibitem{ed} E. Bartz, The 0.25$\mu$m Token Bit Manager Chip for the CMS Pixel Readout, Proceedings of the 11$^{th}$ Workshop on Electronics for LHC and Future Experiments, Heidelberg, Germany
\bibitem{hephy} Designed by M. Pernicka, et al., HEPHY Institute, Vienna, Austria
\bibitem{dorokhov} A. Dorokhov, Performance of Radiation Hard Pixel Sensors for the CMS Experiment, Ph.D. thesis University of Z\"urich, 2006
\end{thebibliography}
\end{document}